\documentclass[12pt]{article}
\usepackage{amsmath}
\usepackage{mathptmx}
\usepackage[mathcal]{euscript}
\usepackage{bm}
\usepackage{graphicx}
\renewcommand{\cal}{\mathcal}

\def \myfiguress #1#2#3#4#5#6#7#8
{\begin{figure}[ht]
    \begin{center}
        \includegraphics[width=#2 \textwidth]{#1.eps}
        \hfill
        \includegraphics[width=#6 \textwidth]{#5.eps}
        \parbox[t]{#4\textwidth}{\caption {#3}\label{#1}}
        \hfill
        \parbox[t]{#8\textwidth}{\caption {#7}\label{#5}}
    \end{center}
\end{figure} }

\def \myfigures #1#2#3#4#5#6#7#8
{\begin{figure}[ht]
    \begin{center}
        \includegraphics[width=#2 \textwidth, angle = -90]{#1.eps}
        \hfill
        \includegraphics[width=#6 \textwidth, angle = -90]{#5.eps}
        \parbox[t]{#4\textwidth}{\caption {#3}\label{#1}}
        \hfill
        \parbox[t]{#8\textwidth}{\caption {#7}\label{#5}}
    \end{center}
\end{figure} }

\def\myfig #1#2#3#4
{\begin{figure}[ht]\begin{center}
\includegraphics[width=#2 \textwidth, angle = 0]{#1.eps}
\vskip -5 mm
\parbox[t]{#4\textwidth}{\caption{#3}\label{#1}}
\end{center}\end{figure}}

\begin{document}
\title{Bianchi type-I string cosmological model in the presence of a
magnetic flux: exact and qualitative solutions}

\author{Bijan Saha
\thanks {E-mail:~~~bijan@jinr.ru;
URL: http://wwwinfo.jinr.ru/\~\,\! bijan/}~ ,
Victor Rikhvitsky
\thanks {E-mail:~~~rqvtsk@jinr.ru}\\
{\small \it Laboratory of Information Technologies}\\
{\small \it Joint Institute for Nuclear Research, Dubna}\\
{\small \it 141980 Dubna, Moscow region, Russia}
\and
Mihai Visinescu
\thanks{E-mail:~~~mvisin@theor1.theory.nipne.ro;
URL: http://www.theory.nipne.ro/\~\,\!mvisin/}\\
{\small \it Department of Theoretical Physics}\\
{\small \it National Institute for Physics and Nuclear Engineering}\\
{\small \it Magurele, P. O. Box MG-6, RO-077125 Bucharest, Romania}}
\date{}

\maketitle

\begin{abstract}

A Bianchi type-I  cosmological model in the
presence of a magnetic flux along a cosmological string
is investigated. The objective of this study is to generate solutions
to the Einstein equations using a few tractable assumptions usually accepted
in the literature. The analytical solutions are supplemented with
numerical and qualitative analysis. In the frame of the present model
the evolution of the Universe and other physical aspects are
discussed.

Pacs: 03.65.Pm; 04.20.Ha

\end{abstract}

            \section{Introduction}

Since the observed Universe is almost homogeneous and isotropic, the
space-time is usually described by a
Friedman-Lamaitre-Robertson-Walker (FLRW) cosmology. Although this
is a good approximation, the recent observations suggest the
necessity of exploring beyond it. Also it is believed that in the
early Universe the FLRW model does not give a correct matter
description. The anomalies found in the cosmic microwave background
(CMB) and the large structure observations stimulated a growing
interest in anisotropic cosmological models of the Universe.

Recently, cosmic strings have  drawn considerable interest among the
re\-sear\-chers for various aspects such as the study of the early
Universe. The presence of cosmic strings in the early Universe could
be explained using grand unified theories. These strings arise during
the phase transition after the Big Bang explosion as the temperature
goes down below some critical point. It is believed that the
existence of strings in the early Universe gives rise to the density
fluctuations which leads to the formation of the galaxies. Also the
cosmic strings have been used in attempts to investigate anisotropic
dark energy component including a coupling between dark energy and a
perfect fluid (dark matter) \cite{KM}. The cosmic string is
characterized by a stress energy tensor and it is coupled to the
gravitational field.

The presence of magnetic fields in galactic and intergalactic spaces
is well established and their importance in astrophysics is generally
acknowledged (see e. g. the reviews \cite {GR,W} and references therein).
In spite of the fact that the present day magnitude of
the magnetic energy is very small in comparison with the matter
density, it might not have been negligible during early stages of the
evolution of the Universe. Any cosmological model which contains
magnetic fields is necessarily anisotropic taking into account that the
magnetic field vector implies a preferred spatial direction.

Among many possible alternatives, the simplest and most theoretically
appealing of anisotropic models are Bianchi type-I (BI). For studying the
effects of aniso\-trop\-ies in the early Universe on the present observed
Universe, BI models have been investigated from different
points of view. In this class of models it is possible to
accommodate the presence of cosmic strings as an example of an
anisotropy of space-times generated by one dimensional topological
defects.

In what follows we shall investigate the evolution of BI
cosmological models in presence of a cosmic string and magnetic
fluid. The paper has the following structure. We shall review the
basic equations of an anisotropic BI model in the presence of a
system of cosmic string and magnetic field. In Section 3 we
introduce a few plausible assumptions and investigate their
outcomes. The objective of this treatment is to generate exact
solutions to the Einstein equations, supplemented with numerical and
qualitative analysis. At the end we shall summarize the results and
outline future prospects.

    \section{Fundamental Equations and general solutions}

The line element of a BI Universe is
\begin{equation}
ds^2 = (dt)^2 - a_1(t)^2 (dx^1)^2 - a_2(t)^2 (dx^2)^2 - a_3(t)^2
(dx^3)^2\,.
\label{BI}
\end{equation}
There are three scale factors $a_i$ $(i=1,2,3)$ which are functions
of time $t$ only and consequently three expansion
rates. In principle all these scale factors could be different and it
is useful to express the mean expansion rate in terms of the average
Hubble rate:
\begin{equation}\label{Hubble}
H = \frac{1}{3}\Bigl(\frac{\dot a_1}{a_1}+\frac{\dot a_2}{a_2}+
\frac{\dot a_3}{a_3}\Bigr)\,,
\end{equation}
where over-dot means differentiation with respect to $t$.

In the absence of a cosmological constant, the
Einstein's gravitational field equation has the form
\begin{subequations}
\label{BID}
\begin{eqnarray}
\frac{\ddot a_2}{a_2} +\frac{\ddot a_3}{a_3} + \frac{\dot a_2}{a_2}\frac{\dot
a_3}{a_3}&=&  \kappa T_{1}^{1}\,,\label{11}\\
\frac{\ddot a_3}{a_3} +\frac{\ddot a_1}{a_1} + \frac{\dot a_3}{a_3}\frac{\dot
a_1}{a_1}&=&  \kappa T_{2}^{2}\,,\label{22}\\
\frac{\ddot a_1}{a_1} +\frac{\ddot a_2}{a_2} + \frac{\dot a_1}{a_1}\frac{\dot
a_2}{a_2}&=&  \kappa T_{3}^{3}\,,\label{33}\\
\frac{\dot a_1}{a_1}\frac{\dot a_2}{a_2} +\frac{\dot
a_2}{a_2}\frac{\dot a_3}{a_3}+\frac{\dot a_3}{a_3}\frac{\dot
a_1}{a_1}&=&  \kappa T_{0}^{0}\,, \label{00}
\end{eqnarray}
\end{subequations}
where $\kappa$ is the gravitational constant. The energy momentum tensor
for a system of cosmic string and magnetic field in a comoving
coordinate is given by
\begin{equation}
T_{\mu}^{\nu} =  \rho u_\mu u^\nu - \lambda x_\mu x^\nu
+ E_\mu^\nu\,, \label{imperfl}
\end{equation}
where $\rho$ is the rest energy density of strings with massive
particles attached to them and can be expressed as $\rho = \rho_{p} +
\lambda$, where $\rho_{p}$ is the rest energy density of the
particles attached to the strings and $\lambda$ is the tension
density of the system of strings \cite{letelier,pradhan,tade}
which may be positive or negative. Here $u_i$ is the four
velocity and $x_i$ is the direction of the string, obeying the
relations
\begin{equation}
u_iu^i = -x_ix^i = 1, \quad u_i x^i = 0\,. \label{velocity}
\end{equation}

In \eqref{imperfl} $E_{\mu\nu}$ is the electromagnetic field given by
Lichnerowich \cite{lich}. In our case the electromagnetic field tensor
$F^{\alpha \beta}$ has only one non-vanishing component, namely
\begin{equation}
F_{23} = h\,,
\label{f23}
\end{equation}
where $h$ is assumed to be constant.
For the electromagnetic field $E_\mu^\nu$ one gets the following non-trivial
components
\begin{equation}
E_0^0 = E_1^1 = - E_2^2 = - E_3^3 = \frac{h^2}
{2 {\bar \mu} a_2^2 a_3^2}\,.
\label{E}
\end{equation}
where $\bar \mu$ is a constant characteristic of the medium and called
the magnetic permeability. Typically $\bar \mu$ differs from unity
only by a few parts in $10^5$ ($\bar \mu > 1$ for paramagnetic
substances and $\bar \mu < 1$ for diamagnetic).

Choosing the string along $x^1$ direction and using comoving
coordinates we have the following components of energy momentum
tensor \cite{ass}:
\begin{subequations}
\label{total}
\begin{eqnarray}
T_{0}^{0} &=&  \rho +
\frac{h^2}{2 {\bar \mu}}\frac{a_1^2}{\tau^2}\,, \label{t00}\\
T_{1}^{1} &=&  \lambda
+\frac{h^2}{2 {\bar \mu}}\frac{a_1^2}{\tau^2}\,, \label{t11}\\
T_{2}^{2} &=&
-\frac{h^2}{2 {\bar \mu}}\frac{a_1^2}{\tau^2}\,, \label{t22}\\
T_{3}^{3} &=&  -\frac{h^2}{2 {\bar \mu}}\frac{a_1^2}{\tau^2}\,,
\label{t33}
\end{eqnarray}
\end{subequations}
where we introduce the volume scale of the BI space-time
\begin{equation}
\tau = a_1 a_2 a_3 \,, \label{taudef}
\end{equation}
namely, $\tau = \sqrt{-g}$ \cite{sahaprd}. It is interesting to note
that the evolution in time of $\tau$ is connected with the Hubble
rate \eqref{Hubbletau}:
\begin{equation}
\frac{\dot \tau}{\tau} = 3 H\,. \label{Hubbletau}
\end{equation}

In view of $T_{2}^{2} = T_{3}^{3}$ from \eqref{22}, \eqref{33} one
finds
\begin{equation}
a_2 = c_3 \, a_3 \ {\rm exp}\Bigl(d \int \frac{dt}{\tau}\Bigr)\,,
\label{b/c}
\end{equation}
with $c_3, d$ some real constants.

Since the metric functions can be
expressed in terms of $\tau$,  let us first derive the equation for
$\tau$. Summation of Einstein equations \eqref{11}, \eqref{22}, \eqref{33}
and 3 times \eqref{00} gives
\begin{equation}
\frac{\ddot \tau}{\tau}= \frac{3}{2}\kappa \Bigl(\rho +
\frac{\lambda}{3} +
 \frac{h^2}{3 {\bar \mu}}\frac{a_1^2}{\tau^2} \Bigr)\,.
\label{dtau1}
\end{equation}

Taking into account the conservation of the energy-momentum tensor,
i.e., $T_{\mu;\nu}^{\nu} = 0$, after a little manipulation of
\eqref{total} one obtains
\begin{equation}
\dot \rho + \frac{\dot \tau}{\tau}\rho - \frac{\dot a_1}{a_1}\lambda
= 0\,. \label{vep}
\end{equation}

\section{Some examples and explicit solutions}

The above equations involve some unknowns and we need some
supplementary
relations between them to have a tractable problem. It is customary to
assume a relation between $\rho$ and $\lambda$ in accordance with
the state equations for strings.
The simplest one is a proportionality relation \cite{letelier}:
\begin{equation}\label{rhoalphalambda}
\rho = \alpha \lambda \,.
\end{equation}
The  most usual choices of the constant $\alpha$ are
\begin{equation}\label{alpha}
\alpha =\left \{
\begin{array}{ll}
1 & \quad {\rm geometric\,\,\,string}\\
1 + \omega  & \quad \omega \ge 0, \quad p \,\,{\rm string\,\,\,or\,\,\,
Takabayasi\,\,\,string}\\
-1  & \quad {\rm Reddy\,\,\,string}\,.
\end{array}
\right.
\end{equation}

In order  to solve the Einstein equations
completely, we need also to impose some additional conditions.
In what follows we shall illustrate the general considerations by two
examples.

\subsection{Case 1: Hubble rate proportional to an eigenvalue of the
shear tensor}

As a first example, we shall follow the condition introduced by Bali
\cite{bali} assuming that the average Hubble rate H \eqref{Hubble} in
the model
is proportional to the eigenvalue $\sigma_1^1$ of the shear tensor
$\sigma_\mu^\nu$. We shall only give the briefest account here and for
further information the reader should consult \cite{SV}.

For the BI space-time we have
\begin{equation}
\sigma_1^1 = - \frac{1}{3}\Bigl(4\frac{\dot a_1}{a_1}+\frac{\dot a_2}{a_2}+
\frac{\dot a_3}{a_3}\Bigr)\,. \label{s11}
\end{equation}
Writing the aforementioned proportionality condition as
\begin{equation}
H =  p \sigma_1^1\,, \label{conbali}
\end{equation}
one comes to the following relation
\begin{equation}
a_1 = c_{23} \bigl(a_2 a_3 \bigr)^q\,,
\label{a=bc}
\end{equation}
where $q = - (p+1)/(4p+1)$ is related to the proportionality constant
$p$ and $c_{23}$ is an integration constant.

In this case  \eqref{vep} takes the form
\begin{equation}\label{vepcase1}
\dot \rho + \bigl(\rho - \frac{q}{q+1}\lambda \bigr)\frac{\dot
\tau}{\tau} = 0\,,
\end{equation}
and  \eqref{dtau1} now reads
\begin{equation}
\ddot \tau= \frac{3}{2}\kappa (\rho + \frac{\lambda}{3}) \tau +
b_1 \tau^{(q-1)/(q+1)}\,,  \label{dtaucase1}
\end{equation}
where $ b_1 = \kappa \frac{h^2}{2{\bar \mu}}c_{23}^{2/(q+1)}$
is another constant.

Let us now study  \eqref{vepcase1} and \eqref{dtaucase1} for
different equations of state. Assuming the relation (\ref{rhoalphalambda})
between the pressure of the perfect fluid $\rho$ and the tension
density $\lambda$  from  \eqref{vepcase1} one finds
\begin{equation}
\frac{\dot \rho}{\rho} = (\frac{q}{\alpha (q+1)} -1) \frac{\dot
\tau}{\tau}\,,
\end{equation}
with the solution
\begin{equation}
\rho = b_2 \tau^{\frac{q}{\alpha (q+1)} -1}\,, \label{rhocase1}
\end{equation}
while the equation for $\tau$ reads
\begin{equation}
\ddot \tau= \frac{3}{2} \kappa b_2 (1 + \frac{1}{3\alpha})
\tau^{\frac{q}{\alpha (q+1)}} + b_1 \tau^{\frac{q-1}{q+1}}\,,
\label{ddtaucase1}
\end{equation}
involving another constant of integration $b_2$.

This equation can be set in the following form
\begin{equation}
\dot \tau = \sqrt{\frac{\kappa b_2 (3\alpha + 1)(q+1)}{q+\alpha(q+1)}
\tau^{1+q/\alpha(q+1)} + \frac{b_1(q+1)}{q} \tau^{2q/(q+1)} +
{b_3}} \label{vel}
\end{equation}
where $b_3$ is an  integration constant.

More details for this model characterized by \eqref{conbali} including
some numerical computations are given in \cite{SV}.

\subsection{Case 2: Constraints on relative shear anisotropy parameters}

The second case is more involved taking into account that
the anisotropy is connected with the values of the shear distortions.
Besides the generalized Hubble parameter \eqref{Hubble}
we consider two relative shear anisotropy parameters defined by
\cite{barrow}:
\begin{subequations}
\begin{eqnarray}
R &=& \frac{1}{H}\bigl(\frac{\dot a_1}{a_1} - \frac{\dot
a_2}{a_2}\bigr)\,, \label{sp1}\\
S &=& \frac{1}{H}\bigl(\frac{\dot a_1}{a_1} - \frac{\dot
a_3}{a_3}\bigr)\,. \label{sp2}
\end{eqnarray}
\end{subequations}
When $R = S = 0$ the universe will be the isotropic flat Friedman
Universe. In this second example we shall assume a deviation from
the Friedman model.

We shall consider three different assumptions regarding the relative
shear ani\-so\-tro\-py parameters. Although these assumptions are
distinct, they entail mild time dependencies for $R$ and $S$. More
exactly the time evolution of the scale factors $a_i$ is assumed to be very
similar to that of the average Hubble rate $H$.

\subsubsection{Assumption 1: $R$ is constant}
Let us assume that $R = r_1$ with $r_1$ a constant.
In view of \eqref{Hubble} one finds,
\begin{equation}
a_1 = c_2 a_2 \tau^{r_1/3}\,. \label{a/b}
\end{equation}
Then, together with  \eqref{taudef}, \eqref{b/c} we find the following
expressions for metric functions
\begin{subequations}
\label{abcb}
\begin{eqnarray}
a_1 &=& (c_2^2 c_3)^{1/3} \tau^{(3+2r_1)/9}
e^{\frac{d}{3}\int\frac{dt}{\tau}}\,, \label{a1b}\\
a_2 &=& (c_2^{-1} c_3)^{1/3} \tau^{(3-r_1)/9}
e^{\frac{d}{3}\int\frac{dt}{\tau}}\,, \label{a2b}\\
a_3 &=& (c_2 c_3^2)^{-1/3} \tau^{(3-r_1)/9}
e^{\frac{-2d}{3}\int\frac{dt}{\tau}}\,. \label{a3b}
\end{eqnarray}
\end{subequations}
In this case  \eqref{vep} takes the form
\begin{equation}
\dot \rho + \bigl(\rho - \frac{3+2r_1}{9}\lambda \bigr)\frac{\dot
\tau}{\tau} - \frac{d}{3\tau}\lambda = 0\,,
 \label{vepcase2}
\end{equation}
whereas, for $\tau$ in this case we have
\begin{equation}
\frac{\ddot \tau}{\tau}= \frac{3}{2}\kappa \Bigl(\rho +
\frac{\lambda}{3} +
 \frac{h^2}{3 {\bar \mu}}(c_2^2c_3)^{2/3}
\tau^{(4r_1-12)/9}e^{\frac{2d}{3}\int\frac{dt}{\tau}} \Bigr)\,.
\label{dtau1barrow}
\end{equation}

Comparing this equation with the corresponding one (\ref{ddtaucase1})
from {\bf Case 1}, it results that in {\bf Case 2} the equation of
evolution for $\tau$ is more intricate. In \cite{SV} we gave some
asymptotic solutions of this equation. Here we give more general
solution to the equation in question.

In what follows we express equations  \eqref{vepcase2} and
\eqref{dtau1barrow} as a new system of differential equations which is
easier to analyze:
\begin{subequations}
\label{system}
\begin{eqnarray}
\dot \tau &=& 3 H \tau\,,\label{hubblesys} \\
\dot T &=& \frac{T}{\tau}\,,\label{aux}\\
\dot H &=& - 3H^2 + \frac{1}{2}\kappa \Bigl( \frac{3 \alpha
+1}{3\alpha} \rho +  \frac{h^2}{3 {\bar
\mu}}(c_2^2c_3)^{\frac{2}{3}}
\tau^{\frac{(4r_1-12)}{9}}T^{\frac{2d}{3}} \Bigr)\,,\label{H} \\
\dot \rho &=& \Bigl(\frac{d}{3 \alpha}\frac{1}{\tau} - \frac{9
\alpha - 3 - 2r_1}{3\alpha} H \Bigr)\rho\,.\label{rhosys}
\end{eqnarray}
\end{subequations}

This system was investigated qualitatively using numerical methods.
To numerically integrate the above system of differential equations with
rational polynomials in the right hand sides we use the following method:
each of the
variables $V = \{ \tau, H, \rho, T\}$ was substituted by $V =
V_s/V_c$ with $V_s^2 + V_c^2 = 1$. Further we find the common
denominator of the right hand sides and moving away it we find the new
system of equation with new parameters. This removing is equivalent
to the substitution of an independent variable.
In the new variable the right hand sides of the differential equations,
as well as the integrable functions themselves, possess finite
variation.
So in the compact region the integral curves happen to be stable if
the initial value is given inside the region. It does not change if
we numerically integrate them in natural parameters (along the
length of the curve).

For exemplification we make the following specific choices: $r_1 =
5.25$, $d = 1.5$, $\kappa = 1$, $\bar{\mu} = 1$, ${\cal J} =1$, $c_3
= 1$, and $c_2 = 1$. We have investigated the system for two
different values of $\alpha$, namely, $\alpha =1$ and $\alpha = -1$.
To plot the graphs\footnote{Figures of this article are in color in
the electronic version.}, it is more convenient to use instead of
the function $f$, $\arctan{f}$ in order to convert their infinite
ranges of variation into finite ones.  The initial condition we give
on the line where $H = 0$ and study the evolution of all the four
functions both through the past and future. In the Figs. \ref{x11}
and \ref{x-11} we plot the evolution of $\tau$ for a positive and
negative $\alpha$, respectively, while in Figs. \ref{x13} and
\ref{x-13} we do the same for energy density. In Figs. \ref{x13D}
and \ref{x-13D}, {\bf 3D} graph of $\tau$, $H$ and $\rho$  has been
illustrated for a positive and negative $\alpha$, respectively for
the following initial values:

\begin{center}
\begin{tabular}{|r|r|r|r|r|r|r|} \hline
\multicolumn{7}{|c|}{initial values} \\\hline $\tau$ & 1.08& 3.53&
0.50& 1.37& 6.28 & 1.78\\\hline $H$ & -1 &
-1&0&0&0&1\\\hline $\rho$ & 1&1&1&1&1&1\\ \hline $T$&.1&.1&.1&.1&.1&.1 \\
\hline
\end{tabular}
\end{center}

\vskip 1 cm

\myfigures{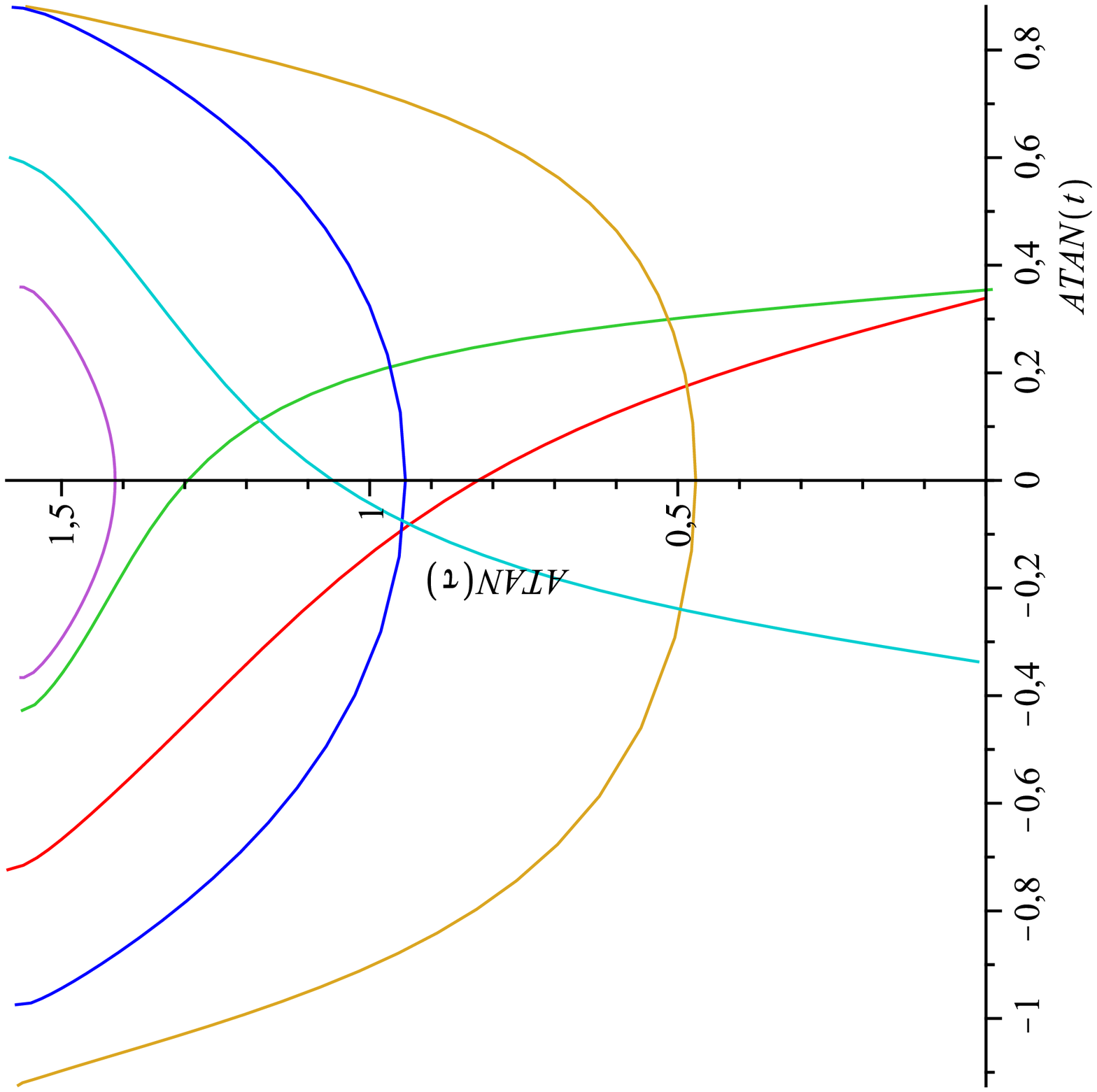}{0.35}{Evolution of the BI universe for a positive
$\alpha$.}{0.35}{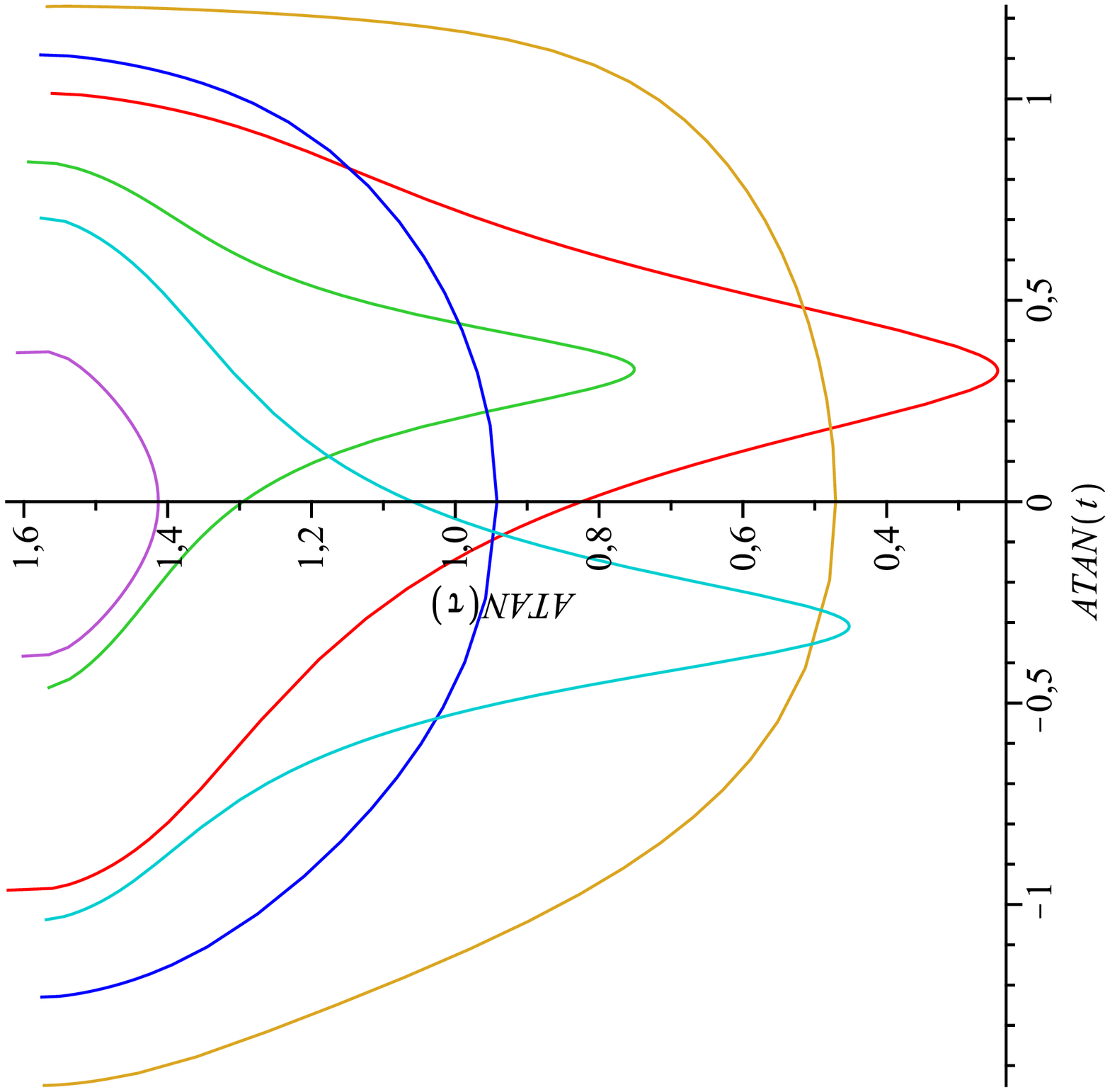} {0.35}{Evolution of the BI universe for a
negative $\alpha$.}{0.35}

\myfigures{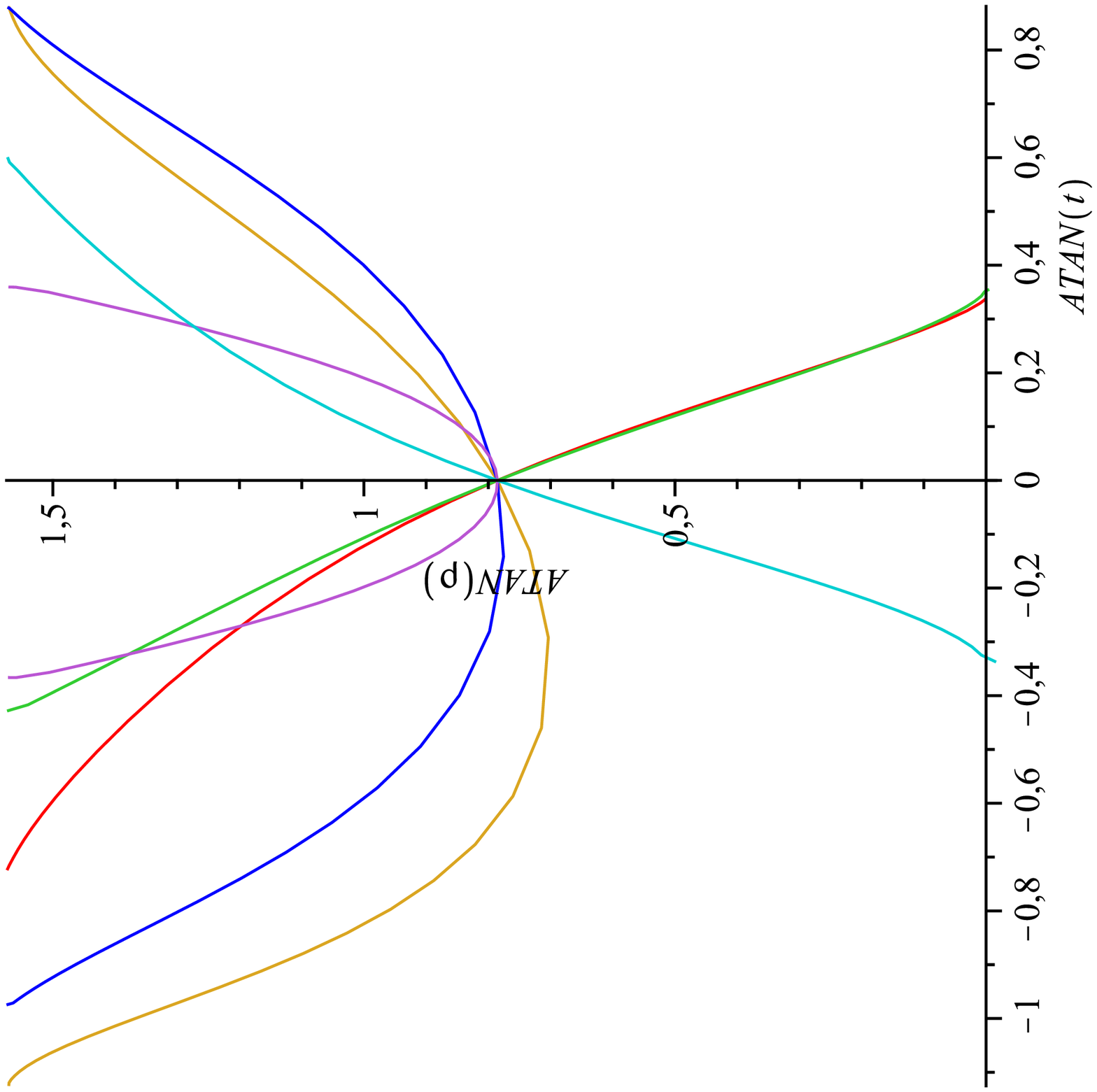}{0.35}{Evolution of the energy density for $\alpha >
0$.}{0.35}{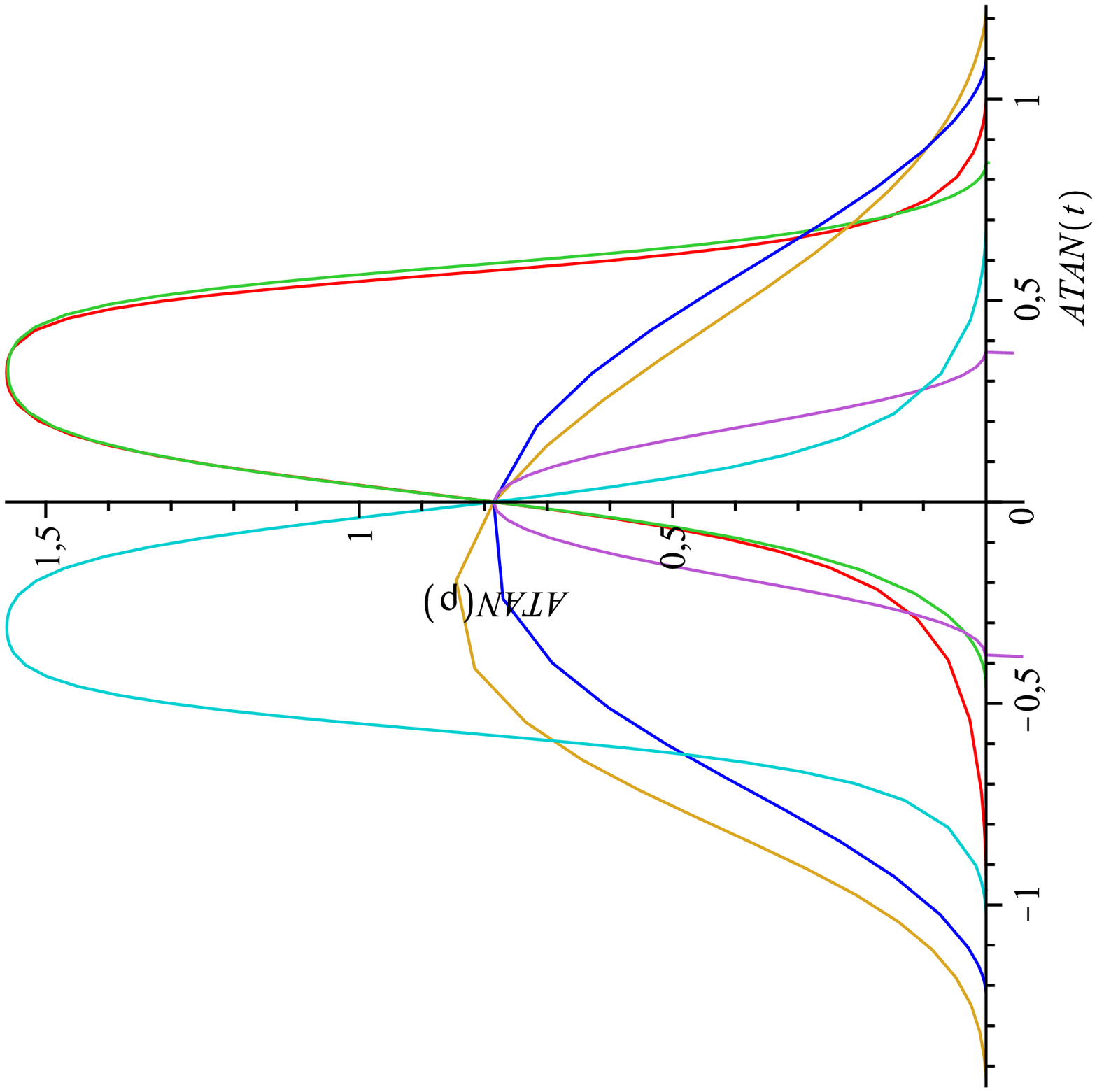} {0.35}{Evolution of the energy density for $\alpha
< 0$.}{0.35}

\myfig{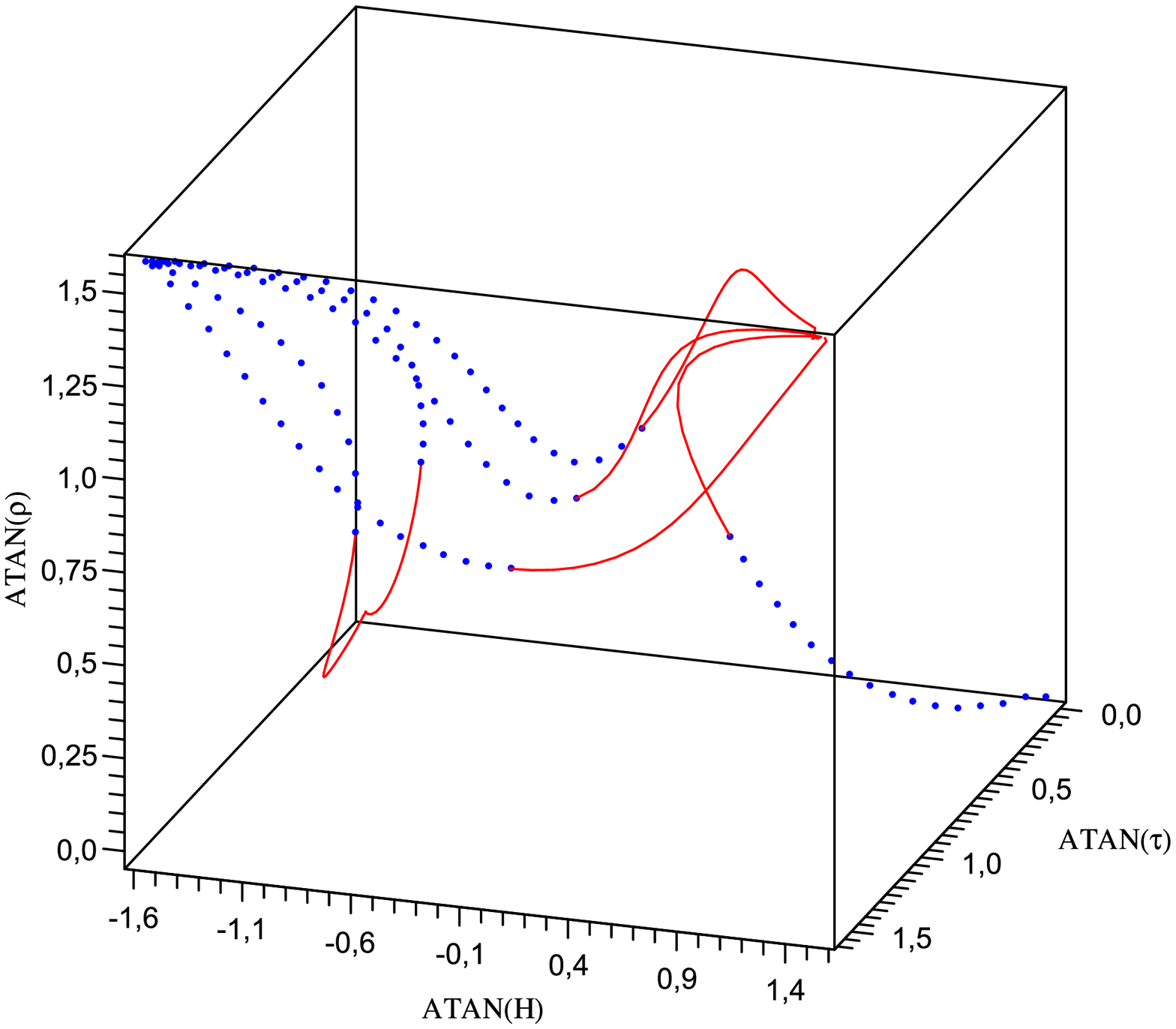}{0.55}{3D view of $\arctan(\rho)$, $\arctan(H)$ and
$\arctan(\tau)$ for $\alpha
> 0$.}{0.85}

\myfig{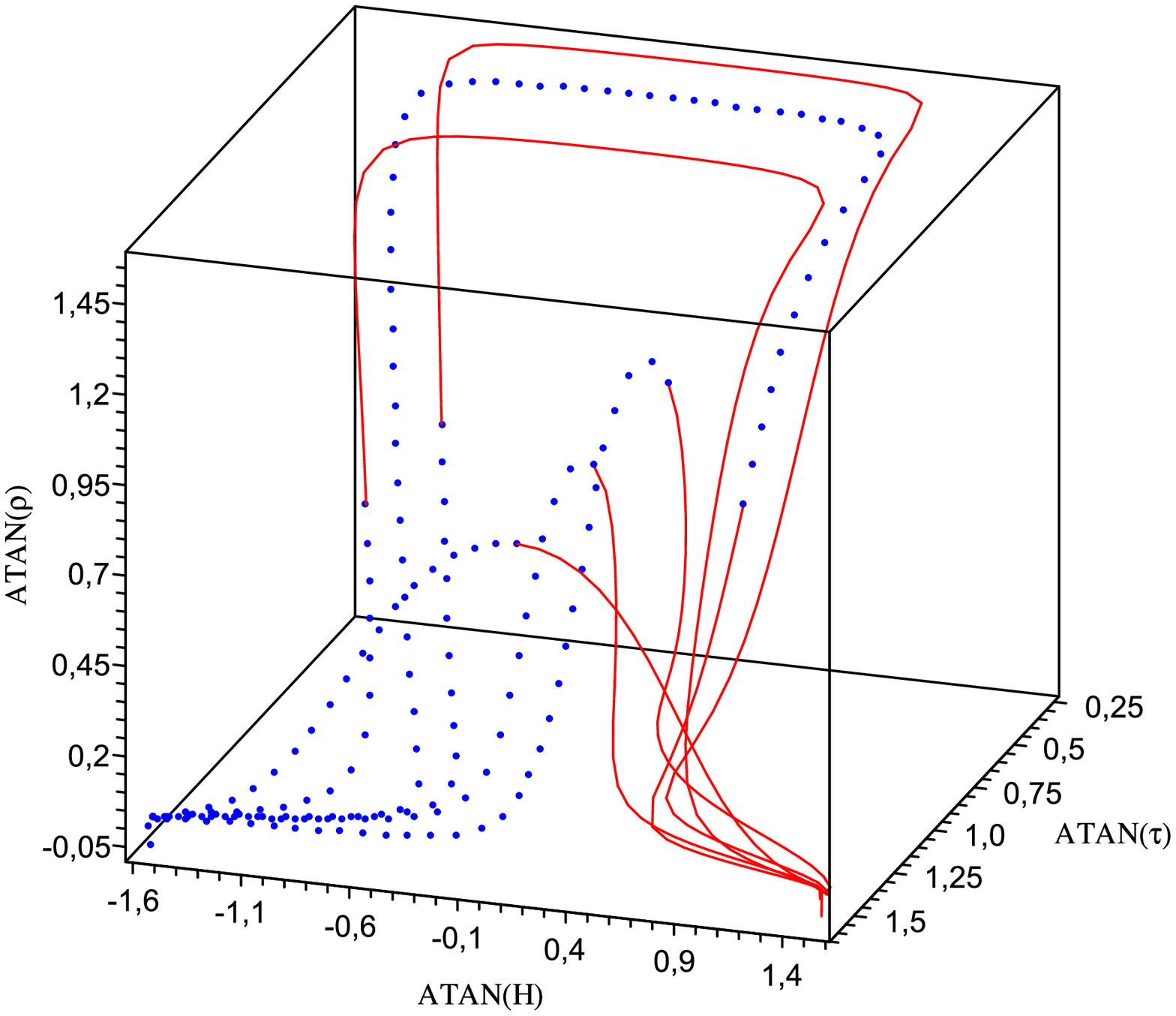}{0.55}{3D view of $\arctan(\rho)$, $\arctan(H)$ and
$\arctan(\tau)$ for $\alpha < 0$.}{0.85}

The system of four time-depended variables $\tau, H, \rho, T$
possesses integral curves in 4D space. The graphs of $\rho(t)$
and $\tau(t)$ are the
projections from the extended space of dimension 4+1 on the plane
with dimension 1+1. Therefore, the projections of integral curves
can intersect. In the figures the integral curves correspond to
different initial conditions with similar other parameters. In the
figures \ref{x13D} and \ref{x-13D} the dot lines leads to the past,
the solid ones to the future, while the point is the one where the
initial condition of integration was given.

There is also the possibility to assume that $S$ is constant and the
analysis develops in the aforesaid manner, practically interchanging
$a_2$ with $a_3$.

\subsubsection{Assumption 2: $R$ and $S$ differ by a constant}
Let us now assume the case when $R - S = r_2$ with $r_2$ a constant.
This assumption leads to
\begin{equation}
a_2 = c_3 a_3 \tau^{r_2/3}\,,  \label{a23}
\end{equation}
which together with \eqref{b/c} gives
\begin{equation}
\tau = \tau_0 + (3d/r_2) t\,. \label{taulin}
\end{equation}
Thus we find $\tau$ to be a linear function of $t$. It should be
noted that analogical results occurs when the BI Universe is filled
with stiff fluid. On the other hand from \eqref{taudef} one finds

\begin{equation}
a_1 = (1/c_3) a_3^{-2} \tau^{1 - r_2/3}\,. \label{a13}
\end{equation}
In view of \eqref{taulin}, \eqref{a13} and \eqref{rhoalphalambda}
from \eqref{dtau1} one finds
\begin{equation}
\rho = - \frac{\alpha h^2}{{\bar \mu} c_3^2 (3 \alpha +
1)}\frac{\tau^{(4r_2 - 6)/3}}{a_3^4}\,. \label{rhoa3}
\end{equation}
Finally, inserting \eqref{rhoa3}, \eqref{a13} into \eqref{vep} we
find
\begin{equation}
a_3 = c_4 \tau^{(\alpha (3 - 2r_2) - 3 + r_2)/6(2\alpha - 1)}\,,
\label{a3fin}
\end{equation}
with $c_4$ another constant.

Thus the Einstein system of equations has been completely solved .

\subsubsection{Assumption 3: $R$ and $S$ are proportional}
Let us now assume the case when $R = r_3 \, S$
with $r_3$ being some constant. This leads to the relation

\begin{equation}
\frac{a_1^{1-r_3}a_3^{r_3}}{a_2} = c_{13},\label{a123rel}
\end{equation}
with $c_{13}$ being some arbitrary constant. This relation together with
\eqref{b/c} and \eqref{taudef} gives
\begin{subequations}
\begin{eqnarray}
a_1 &=& c_3^{(r_3+1)/3(1-r_3)} c_{13}^{2/3(1-r_3)} \tau^{1/3}
e^{\frac{d}{3}\frac{r_3+1}{1-r_3} \int \frac{dt}{\tau}}\,, \label{a1n}\\
a_2 &=& c_3^{(1-2r_3)/3(1-r_3)} c_{13}^{-1/3(1-r_3)} \tau^{1/3}
e^{\frac{d}{3}\frac{1-2r_3}{1-r_3} \int \frac{dt}{\tau}}\,, \label{a2n}\\
a_3 &=& c_3^{(r_3-2)/3(1-r_3)} c_{13}^{-1/3(1-r_3)} \tau^{1/3}
e^{\frac{d}{3}\frac{r_3-2}{1-r_3} \int \frac{dt}{\tau}}\,. \label{a3n}
\end{eqnarray}
\end{subequations}

The system analogue to \eqref{system} is:
\begin{subequations}
\label{system01}
\begin{eqnarray}
\dot \tau &=& 3 H \tau\,,\label{hubblesys01} \\
\dot T &=& \frac{T}{\tau}\,,\label{aux01}\\
\dot H &=& - 3H^2 + \frac{1}{2}\kappa \Bigl(\frac{3\alpha +
1}{3\alpha} \rho + \frac{h^2}{3 {\bar
\mu}}c_3^{\frac{2}{3}\frac{r_3+1}{1-r_3}}c_{13}^{\frac{4}{3(1-r_3)}}
\tau^{4/3}T^{\frac{2d}{3}\frac{1+r_3}{1-r_3}} \Bigr),\label{H01} \\
\dot \rho &=& \Bigl(\frac{d}{3\alpha} \frac{1+r_3}{1-r_3}\frac{1}{\tau}
+ \frac{1 - 3\alpha}{\alpha} H \Bigr)\rho\,.\label{rhosys01}
\end{eqnarray}
\end{subequations}

This system is quite similar to \eqref{system} and choosing
suitable parameters we find solutions which look as the ones
illustrated in the previous graphs.

\section{Summary and outlook}
We have offered an investigation of an anisotropic cosmological BI
model. Having in mind the complexity of the model we used some
tractable assumptions regarding the parameters entering the model.
For different assumptions regarding the shear anisotropy parameters
\eqref{sp1} and \eqref{sp2} we
get interesting models deserving the study.
The analytical results are supplemented with numerical and qualitative
analysis describing the evolution of a BI Universe for different values
of the parameters.

In our further studies \cite{BVMnew} we should like to see how the
model isotropises at late times. Any realistic model must lead to
isotropisation necessary for compatibility with standard
cosmological models at late times and in agreement with current
observations. Also it is important to know how stable is the model
to perturbations of the parameters.

\subsection*{Acknowledgments}
The authors gratefully acknowledge the support from the
joint Romanian-LIT, JINR,
Dubna Research Project, theme no. 05-6-1060-2005/2010. The work
of M. V. is supported in part by CNCSIS Programs, Romania.


\end{document}